\documentclass{aa}
\input{psfig}

\def\p{\partial}
\def\e{{\rm e}}
\def\d{{\rm d}}
\def\ie{i.e. }
\def\eg{e.g. }
\def\etal{et al. }

 \def\es{{\rm ~ergs~s}^{-1}}
\def\kms{{\rm ~km~s}^{-1}}

\def\XX{{\rm ad}}
\def\Var{{\rm Var}\;}
\newlength{\largeur}
\newlength{\saut}
\def\marge#1{
\setlength{\largeur}{\columnwidth}
\addtolength{\largeur}{-#1}
\setlength{\saut}{0.5\largeur}\hspace*{\saut}} \def\picture #1 by #2 (#3){
\marge{#1} \vbox to #2{
\hrule width #1 height 0pt depth 0pt
\vfill
\special{picture #3}}}
\begin{document}

\thesaurus{9 (06.15.1, 06.01.02, 06.06.3, 06.13.1, 03.13.1, 03.13.6)} 

\title{Are solar granules the only source of acoustic oscillations ?} 

\author{T. Foglizzo}

\offprints{foglizzo@cea.fr}

\institute {
Service d'Astrophysique, CEA/DSM/DAPNIA, CE-Saclay, 91191 
Gif-sur-Yvette, France}

\date{}

\maketitle

%\markboth{}{}

\begin{abstract}
The excitation mechanism of low degree acoustic modes is investigated 
through the analysis of the stochastic time variations of their energy.
The correlation between the energies of two different modes is interpreted 
as the signature of the occurrence rate of their excitation source. 
The different correlations determined by Foglizzo \etal (1998) 
constrain the physical properties of an hypothetical source of excitation, 
which would act in addition to the classical excitation by the turbulent 
convection. Particular attention is drawn to the effect of
coronal mass ejections. The variability of their occurrence rate with the 
solar cycle could account for the variation of the correlation between 
IPHIR and GOLF data. Such an interpretation would suggest that the mean
correlation between low degree acoustic modes is at least $0.2\%$ at solar 
minimum. 


\end{abstract}

\section{Introduction}
The theoretical mechanism of excitation of solar acoustic modes by 
the turbulent
convection is well documented
(Goldreich \& Keeley 1977; Goldreich \& Kumar 1988; 
Kumar \& Goldreich 1989; Balmforth 1992a,b,c; Goldreich \etal 1994). 
High frequency 
oscillations were interpreted in different ways (Kumar \& Lu 
1991; Goode \etal 1992; Restaino \etal 1993;
Vorontsov \etal 1998), in order to address the question of the 
depth of the excitation source.
Local observations led Rimmele \etal (1995) and Espagnet \etal (1996) 
to identify the excitation of 5-minute 
oscillations with acoustic events occurring in the downflowing 
intergranular regions, rather than overshooting granules.\\ 
Considering granules of $1000$ km diameter renewed on a time scale 
comparable with their turnover time ($8$ min), there are about 
$5\times 10^{9}$ excitations by solar granules per damping time 
($3.7$ days). Even if the number of efficient excitations is smaller by a 
factor 100 (Brown 1991), each low degree mode is stochastically 
excited so many times per damping time, that modes with different
frequencies are expected to have uncorrelated energies.
In particular, the fitting procedures used to determine the 
frequency, linewidth, and splitting of p modes often rely on the
statistical independence of neighbouring modes 
(\eg Appourchaux \etal 1998).\\
Baudin \etal (1996), however, noticed a possible correlation between 
low degree
p modes in IPHIR data. By measuring the mean correlation of p modes in 
IPHIR and GOLF data, and determining their statistical significance, 
Foglizzo \etal (1998) (hereafter F98) suggested
the existence of an additional source of excitation during the IPHIR 
period (end of 1988).\\
The goal of this study is to use the observed correlations as a 
constraint on the physical properties of this hypothetical additional 
source. The theoretical relationship between the occurrence rate
of this source, its contribution to the energy of low degree p modes, 
and their correlation coefficient, is established in Sect.~2. 
Available observational constraints are reviewed in Sect.~3. 
The possible role of comets, X-ray flares and coronal mass
ejections (CMEs) is discussed in Sect.~4.

\section{Correlation coefficient between two modes excited by a single 
mechanism\label{seccorr}}

\subsection{The energy of a mode seen as a random walk\label{secrw}} 

Let us consider impulsive excitations, distributed over the 
solar surface. The exciting events, indexed by $k$, occur at the random 
poissonian time $t_{k}$. They correspond to a radial velocity perturbation
$v_{k}({\bf r})$ localized in a cone of angle $\alpha_{k}$ around the 
random direction ($\theta_{k},\phi_{k}$), with a radial extension 
$\delta r_{k}$. 
$v_{k}({\bf r})$ is described by an amplitude 
$v_{k}$ and dimensionless shape functions $g_{k}$ and $h_{k}$ as 
follows: 
\begin{eqnarray}
v_{k}({\bf r})&\equiv& g_{k}(\alpha) h_{k}(r)v_{k},\label{defv1}\\ 
h_{k}(r)&\sim& 1\;\;{\rm for}\;\;R_{\star}>r>R_{\star}-\delta r_{k},
\label{defv2}\\ 
g_{k}(\alpha)&\sim& 1\;\;
{\rm for}\;\;\alpha<\alpha_{k}. \label{defv3}
\end{eqnarray}
$\alpha(\theta,\theta_{k},\phi-\phi_{k})\ge 0$ is the angle made by the 
direction $\theta,\phi$ with 
the direction $\theta_{k},\phi_{k}$. Let $M_{k}$ be the mass of gas 
in the volume defined by the functions $g_{k}$ and $h_{k}$, so that 
the kinetic energy of the perturbation is ${\cal E}_{k}=M_{k}v_{k}^{2}/2$.
The mean interpulse time is
defined as $\Delta t\equiv <t_{k+1}-t_{k}>$.\\ 
We assume that each impulsive event triggers free oscillations of the set 
of p modes. Each perturbation is projected onto the basis of eigenvectors. 
The damping time $\tau_{d}$ depends in principle on the mode 
considered, but is nearly constant ($\sim~3.7$ days) for the modes 
considered by F98, in the plateau region between
$2.5$mHz and $3.5$mHz (linewidth of $ 1 \mu$Hz).\\
In appendix~A it is shown that the total energy of the oscillations 
associated to the frequency $\omega_{nl}+m\Omega$ is: 
\begin{eqnarray}
E_{nlm}^{+}(t)&=&\left|\sum_{t_{k}<t} 
\e^{-{t-t_{k}\over\tau_{d}}} a_{nlm}^{k}
\e^{i\Phi^{k}_{nlm}}\right|^{2},\label{enerstoc}\\ 
a_{nlm}^{k}&\equiv& 
g_{lm}^{k}(\theta_{k})h_{nl}^{k}M_{k}^{1\over2}v_{k},\\
\Phi^{k}_{nlm}&\equiv& (\omega_{nl}+m\Omega)t_{k}+m\phi_{k} 
\label{defphi}.
\end{eqnarray}
where the real function $g_{l,m}^{k}=g_{l,-m}^{k}$ depends
on the angular shape of the excitation projected onto the spherical 
harmonics, and $h_{nl}^{k}$ depends on the radial shape of the excitation 
projected onto the
radial part $v_{nl}^{r}(r)$ of the eigenfunction:
\begin{eqnarray}
g_{lm}^{k}(\theta_{k})&\equiv &q_{lm}\int_{0}^{{\alpha_{k}\over 
2}}\d\phi\int_{0}^{\pi}\d(\cos\theta) 
P_{l}^{|m|}(\cos\theta)\nonumber\\
&\times&\cos(m\phi)g_{k}(\theta,\theta_{k},\phi),\label{glm}\\ 
h_{nl}^{k}&\equiv&{1\over M_{k}^{1\over2}}
\int_{R_{\star}-\delta r_{k}}^{R_{\star}} 
\rho r^{2}v_{nl}^{r}(r)h_{k}(r)\d r. 
\label{alphanlm}
\end{eqnarray}
The relationship between the spherical harmonics and the Legendre 
associated functions $P_{l}^{m}$ through the constant $q_{lm}$ is 
recalled in Eqs.~(\ref{spha1})-(\ref{spha2}).
The exponential damping in Eq.~(\ref{enerstoc}) can be schematized as 
a term selecting the finite set of excitations which occurred within one 
damping time before the time $t$:
\begin{equation}
E_{nlm}^{+}(t)\sim \left | \sum_{0<t-t_{k}<\tau_{d}} 
a_{nlm}^{k}\e^{i\Phi^{k}_{nlm}}\right |^{2}.\label{enlmap} 
\end{equation}
The series of phases $\omega_{nl}t_{k}$ and $\phi_{k}$ are 
independent random variables uniformly distributed in the interval 
$[0,2\pi]$.
The series $\Phi^{k}_{nlm}$ defined by Eq.~(\ref{defphi}) is 
therefore also uniformly distributed in
$[0,2\pi]$. Ignoring the academic case
where the ratio of the frequencies is a simple rational number, 
$\Phi^{k}_{nlm}$ and $\Phi^{k}_{n'l'm'}$ can be considered independent 
for two different realistic modes.\\ 
As a consequence, the energy of the wave is 
interpreted as the squared length of a random walk in 
the complex plane. Each step of this random walk is defined
by an amplitude $a_{nlm}^{k}$, and a phase 
$\Phi^{k}_{nlm}$ (negative values of the amplitude can be converted into 
an increment of the phase). The number of steps $N$ is the number of 
excitations, over the whole surface of
the sun, within one damping time:
\begin{equation}
N\equiv{\tau_{d}\over\Delta t}.
\end{equation}
The central limit theorem ensures that the real and imaginary parts 
of the complex sum (\ref{enlmap}) converge towards independent normal 
distributions if $N$ is large enough, resulting in an exponential 
distribution of energy (see also Kumar \etal 1988).\\ 
According to this random walk interpretation, two different modes 
excited by the
same series of events must have correlated energies. This correlation 
should approach $100\%$ if the number of steps of this
random walk is small, \ie if the interpulse mean time $\Delta t$ is 
longer than the damping time $\tau_{d}$.\\ 
The independence of the phases $\Phi^{k}_{nlm}$ and 
$\Phi^{k}_{n'l'm'}$, however, makes the correlation decrease to zero 
when the number of excitations increases.\\
In particular, the random walks associated with the two components 
$\omega_{nl}\pm m\Omega$ of a mode $n,l$ have the same series of 
amplitudes, but have two independent series of phases.
This leads us to expect that waves travelling in opposite directions 
do not have the same instantaneous total energy (except on
average), although they are excited by the same events. 

\subsection{Theoretical correlation between the energies of two modes
\label{seccor}} 

The mean energy $<E_{nlm}>$ of the mode $nlm$ is directly proportional 
to the 
mean energy $<e_{nlm}>$ received from each excitation (appendix~B):
\begin{equation}
<E_{nlm}>= {N\over2}<e_{nlm}>.\label{azerty}
\end{equation}
The correlation ${\cal C}_{nlm}^{n'l'm'}$
between two modes $nlm$, $n'l'm'$ excited by the same 
source described by Eq.~(\ref{enerstoc}) is derived in appendix~B:
\begin{eqnarray}
{\cal C}_{nlm}^{n'l'm'}={<e_{nlm}e_{n'l'm'}>\over
<e_{nlm}^{2}>^{1\over2}<e_{n'l'm'}^{2}>^{1\over2}}\nonumber\\
\times\left(
1+N{<e_{nlm}>^{2}\over<e_{nlm}^{2}>}\right)^{-{1\over2}}
\left(
1+N{<e_{n'l'm'}>^{2}\over<e_{n'l'm'}^{2}>}\right)^{-{1\over2}}.
\label{corgen1}
\end{eqnarray}

\subsection{Latitudinal distribution and sizes of the 
excitations\label{size}}

The ratio $<e_{nlm}^{2}>/<e_{nlm}>^{2}$, related to the spread
of the distribution, is called kurtosis (it is sometimes defined with an 
additional constant $-3$ which we omit here). 
\begin{equation}
{<e_{nlm}^{2}>\over<e_{nlm}>^{2}}= 
{<g_{lm}^{4}h_{nl}^{4}{\cal E}^{2}>
\over <g_{lm}^{2}h_{nl}^{2}{\cal E}>^{2}}.
\end{equation}
The kurtosis of $e_{nlm}$ is 
partly produced by the kurtosis of ${\cal E}_{k}$ (independent of $nlm$), 
but also by the projection of the spatial distribution of excitations 
($\theta_{k},\alpha_{k},\delta r_{k}$) onto the eigenfunctions $nlm$. \\
The random variations of the size $\alpha_{k},\delta r_{k}$ of the 
excitation play a negligible role as long as it is smaller than the 
wavelength of the mode $nlm$. For the modes $l=0$ and $l=1$ analysed by F98, 
we restrict ourselves to perturbations such that
$\alpha_{k}<\pi$, and $\delta r_{k}$ is smaller than the depth of the first
radial node of the eigenfunction $v_{nl}$, and thus neglect the random 
variations of $\alpha_{k},\delta r_{k}$.\\
Let us estimate the kurtosis of the distribution 
$g_{lm}(\theta_{k})$, due to the projection of the latitudinal 
distribution of excitations on the mode $nlm$.
From Eq.~(\ref{glm}), the function $g_{lm}(\theta_{k})$ is 
independent of $\theta_{k}$ for the radial mode $l=0$:
\begin{equation}
{<g_{0,0}^{4}>\over<g_{0,0}^{2}>^{2}}=1.
\end{equation}
By contrast, the mode $l=1,m=\pm1$ is more excited by equatorial events 
than polar ones. For small scale excitations ($\alpha_{k}\equiv 0$) 
distributed uniformly over the sphere, Eq.~(\ref{glm})
implies:
\begin{equation}
{<g_{1,\pm1}^{4}>\over<g_{1,\pm1}^{2}>^{2}}={6\over5}.\label{rat11}
\end{equation}
The kurtosis of $g^{2}_{1,\pm1}$ should reach a value even closer to 
unity if the excitations are distributed in an equatorial region, 
like CMEs at solar minimum or big flares.\\
Let us approximate the kurtosis of $e_{nlm}$ as the product of the 
kurtosis of ${\cal E}_{k}$ by the kurtosis of $g^{2}_{lm}(\theta_{k})$.
\begin{equation}
{<e_{nlm}^{2}>\over<e_{nlm}>^{2}}\sim 
{<{\cal E}^{2}>\over <{\cal E}>^{2}}\times{<g_{lm}^{4}>
\over <g_{lm}^{2}>^{2}}.
\end{equation}
Using this approximation in Eq.~(\ref{corgen1}), the correlation between 
low degree modes $l=0,m=0$ and $l=1,m=\pm1$ excited in the random 
direction $(\theta_{k},\phi_{k})$ by a small scale excitation 
($\alpha_{k}\equiv 0$) is:
\begin{eqnarray}
{\cal C}_{n,0,0}^{n',0,0}&=&\left(1+N_{\rm eff}\right)^{-1},
\label{c00}\\
{\cal C}_{n,0,0}^{n',1,\pm1}&=&\left({5\over6}\right)^{1\over2}
\left(1+N_{\rm eff}\right)^{-{1\over2}}
\left(1+{5N_{\rm eff}\over6}\right)^{-{1\over2}},
\label{c01}\\
{\cal C}_{n,1,\pm1}^{n',1,\pm1}&=&
\left(1+{5N_{\rm eff}\over6}\right)^{-1},
\label{c11}
\end{eqnarray}
where we have incorporated the kurtosis of ${\cal E}_{k}$ 
into the definition of the effective number $N_{\rm eff}$ 
of excitations per damping time: 
\begin{eqnarray}
\beta&\equiv&{<{\cal E}^{2}>\over <{\cal E}>^{2}}\ge1,\\
N_{\rm eff}&\equiv& {N\over\beta}.
\end{eqnarray}
\begin{figure}
\psfig{file=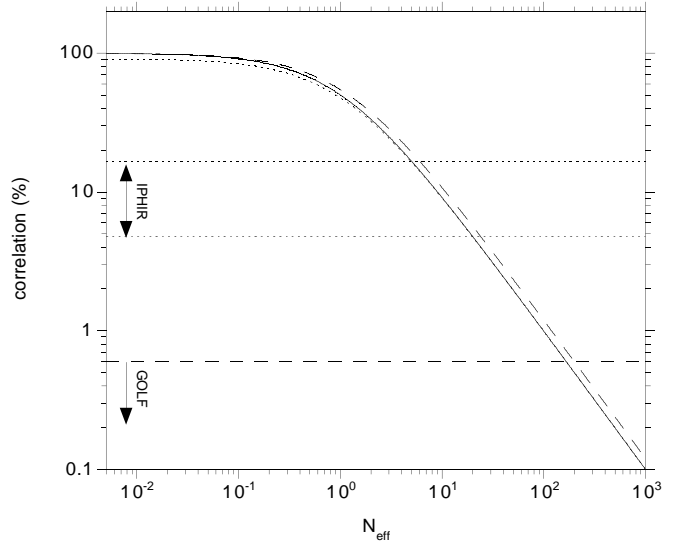,width=\columnwidth} 
\caption[]{Theoretical correlation between two
modes $l=0$ (full line), $l=0$ and $l=1,m=1$ (dotted line), and
$l=1,m=\pm1$ (dashed line), as a function of
the number of effective excitations per damping time 
($\alpha_{k}\equiv 0$). 
Horizontal lines indicate the measured correlation in 
IPHIR data and the upper bound found in GOLF data.} 
\label{fitnum}
\end{figure}
The correlation displayed in Fig.~\ref{fitnum} appears to be hardly 
sensitive to the geometrical content of the function $g_{lm}$, 
since all curves are similar within $10\% $. Error bars of the 
observationally measured correlations being typically larger than $0.1$, 
Eq.~(\ref{c00}) shall be considered accurate enough for 
the purpose of this study:
\begin{equation}
{\cal C}_{n,l,m}^{n',l',m'}\sim\left(1+N_{\rm eff}\right)^{-1}.
\label{fitcor1}
\end{equation}
This is equivalent to approximating the kurtosis of $e_{nlm}$ by the 
kurtosis of ${\cal E}$.

\subsection{Case of two excitation mechanisms superimposed\label{secsup}} 

F98 used a one parameter model of two sources of excitation, such
that each mode contains a fraction $\lambda$ of energy common to all the 
modes. They assumed that both sources produce exponential 
distributions of energy. This simplification led to the simple 
relation $\lambda={\cal C}^{1/2}$. This simplification, however, does 
not apply to the case of a correlation produced by rare impulsive events. 
If $N\ll1$, 
the energy is damped to zero except during isolated pulses of duration 
$\sim\tau_{\rm d}/2$. The distribution $E$ thus follows
a Bernoulli statistics, with ${\rm Var}(E)/<E>^{2}\gg1$ (Eq.~\ref{vaava}) 
instead of $1$ for an exponential distribution. 
The occurrence rate of the excitation mechanism therefore appears to 
be a key parameter which cannot be neglected. Let us consider two 
excitations mechanisms acting simultaneously on two modes $n,l,m$ and 
$n',l',m'$: 
\par(i) excitation by granules, with a high occurrence rate ($N_{1}\gg1$).
\par(ii) excitation by additional events with a total acoustic 
energy ${\cal E}_{\XX}^{k}$, occurring on average $N_\XX$ 
times per damping time, and contributing to a fraction $\lambda_{nlm}$ 
of the power of the mode $nlm$.\\
We show in appendix~C (Eq.~\ref{corgenlamb}) that the correlation between 
the energies of two oscillators excited by these sources is the same as in 
Eq.~(\ref{corgen1}), 
but multiplying the kurtosis of $e_{nlm}$ by $\lambda_{nlm}^{2}$. 
The latitudinal distribution and sizes of 
the excitations play a relatively small role according to Sect.~\ref{size}.
The effective number $N_{\rm eff}$ of excitations 
per damping time depends on the kurtosis $\beta_{\XX}$ of 
${\cal E}_{\XX}$:
\begin{eqnarray}
N_{\rm eff}\equiv {N_{\XX}\over\beta_{\XX}}.
\end{eqnarray}
Making the additional assumption that the 
fraction $\lambda_{nlm}=\lambda$ varies little among the modes 
considered by F98. The mean correlation between these modes is:
\begin{eqnarray}
{\cal C}\sim
\left(1+{N_{\rm eff}\over\lambda^{2}}\right)^{-1}.
\label{coraprox}
\end{eqnarray}
A significant correlation can thus be produced by a source representing 
only a small fraction $\lambda$ of the total power of each mode if 
$N_{\rm eff}$ is small enough. \\
Interpreting the observed correlation ${\cal C}$ as a consequence of 
an additional source of excitation therefore requires it 
to contribute to a fraction $\lambda$ of the total power, deduced from 
Eq.~(\ref{coraprox}):
\begin{equation}
\lambda\sim
\left\lbrace{{\cal C}\;N_{\rm eff}\over1-{\cal 
C}}\right\rbrace^{1\over2}. \label{apllic}
\end{equation}
According to Eq.~(\ref{azerty}), the fraction $\lambda$ of power 
coming from the additional source is related to the mean acoustic energy 
input $<e_\XX>$ into the mode $nlm$:
\begin{equation}
\lambda={N_\XX\over2}{<e_\XX>\over <E>}. \label{frala}
\end{equation}
Eq.~(\ref{apllic}) becomes:
\begin{eqnarray}
<e_\XX>&=&
{2\over \beta_{\XX}}
{{\cal C}\over1-{\cal C}}
{<E>\over \lambda},\label{elam}\\ 
&=&{2\over\beta_{\XX}^{1\over2}} 
\left\lbrace{{\cal C}\over1-{\cal C}}\right\rbrace^{1\over2} 
{<E>\over N_\XX^{1\over2}}.
\label{apllic11}
\end{eqnarray}
Let us assume that the occurrence rate $N_\XX$ varies from $N_{\rm min}$ 
to $N_{\rm max}$ with the solar cycle, while the properties of the 
exciting events ($<e_\XX>,\beta_\XX$) remain unchanged. We use 
Eq.~(\ref{apllic11}) to express the relationship between the
minimum and maximum correlations 
${\cal C}_{\rm min},{\cal C}_{\rm max}$ :
\begin{equation}
{\cal C}_{\rm min}=\left\lbrace
1+{N_{\rm max}\over N_{\rm min}}
{1-{\cal C}_{\rm max}\over{\cal C}_{\rm max}}
\left({<E>_{\rm min}\over<E>_{\rm max}}\right)^{2}
\right\rbrace^{-1}.\label{minmax}
\end{equation}

\subsection{Efficiency of the generation of acoustic waves\label{effic}}

Since the distribution of acoustic energy ${\cal E}_\XX$ produced by the 
additional source is not directly observable, 
we are led to assume that it resembles the observed distribution of total 
energy ${\cal E}_{\rm T}$ of the additional source.
Let $p_{\rm T}({\cal E})$ be the density of probability of the source of 
energy ${\cal E}_{\rm min}\le{\cal E}_{\rm T}\le{\cal E}_{\rm max}$,
occurring $N_{\rm T}$ times per damping time. Let 
us assume that the excitation of low degree modes is efficient only
in the range ${\cal E}_{1}\le{\cal E}_{\rm T}\le{\cal E}_{2}$, inside 
which the efficiency $f_{\XX}\equiv {\cal E}_\XX/{\cal E}_{\rm T}$ is 
constant:
\begin{eqnarray}
{\cal E}_\XX &=& f_{\XX} {\cal E}_{\rm T}\;\;
{\rm for }\;{\cal E}_{1}\le{\cal E}_{\rm T}\le{\cal E}_{2},\\
{\cal E}_\XX &=& 0\;\;{\rm otherwise}.
\end{eqnarray}
In appendix~C, Eq.~(\ref{apllic11}) is rewritten as the fraction of 
the mean acoustic energy $<{\cal E}_\XX>$ which must be injected into 
each mode $nlm$ in order to produce the observed correlation:
\begin{equation}
{<e_\XX>\over<{\cal E}_\XX>}
={2\over f_{\XX}
\left(
\int_{{\cal E}_{1}}^{{\cal E}_{2}}{\cal E}^{2}p_{\rm T}({\cal E})
\d {\cal E}
\right)^{1\over2}} 
\left\lbrace{{\cal C}\over1-{\cal C}}\right\rbrace^{1\over2} 
{<E>\over N_{\rm T}^{1\over2}}.\label{filtre}
\end{equation}
This fraction is therefore minimal when the range of efficient 
excitations ${\cal E}_{1},{\cal E}_{2}$ contains the range of energies 
where the product ${\cal E}_{\rm T}^{2}p({\cal E}_{\rm T})$ is maximal. 
Using in Eq.~(\ref{apllic11}) the occurrence rate $N_{\rm T}$ and kurtosis 
$\beta_{\rm T}$ of the distribution of total energy ${\cal E}_{\rm T}$ 
instead of the distribution of acoustic energy
${\cal E}_{\XX}$ leads to a lower bound of the ratio 
${<e_\XX>/<{\cal E}_\XX>}$:
\begin{equation}
{<e_\XX>\over<{\cal E}_\XX>}\ge
{1\over f_{\XX}}{2\over N_{\rm T}^{1\over2}\beta_{\rm T}^{1\over2}} 
\left\lbrace{{\cal C}\over1-{\cal C}}\right\rbrace^{1\over2} 
{<E>\over <{\cal E}_{\rm T}>}.
\label{apllic22}
\end{equation}
For a given excitation mechanism one can estimate the number
${\cal N}$ of p modes with a wavelength longer than 
the size of each exciting event, which receive a 
comparable amount of energy from the source. Considering that the 
low degree modes in F98 belong to this set,
\begin{equation}
<{\cal E}_{\XX}>=\sum_{nlm} <e_{nlm}>\ge{\cal N}<e_\XX>.\label{calN}
\end{equation}
From Eqs.~(\ref{apllic22})-(\ref{calN}) we deduce a constraint on 
observable quantities, which shall be useful in Sect.~\ref{fcme} in 
order to discriminate between possible excitation mechanisms:
\begin{equation}
{2\over N_{\rm T}^{1\over2}\beta_{\rm T}^{1\over2}} 
\left\lbrace{{\cal C}\over1-{\cal C}}\right\rbrace^{1\over2} 
{<E>\over <{\cal E}_{\rm T}>}\le {f_{\XX}\over{\cal N}}<{1\over{\cal N}}.
\label{apllic33}
\end{equation}

\section{Observational constraints}

\subsection{Exponential distribution of individual p modes 
energy\label{dis1}}

The distribution of energy of low degree p modes agrees reasonably well 
with an exponential distribution (Chaplin \etal 1995, 1997
for BiSON data, F98 for IPHIR and GOLF data). Chaplin \etal (1995, 1997), 
however, 
noticed significant deviations in the high energy tail of the distribution, 
which could be due to an additional excitation mechanism. 
They estimated that the probability that such 
deviations occur by chance during the period of observation
is only $0.1\% $.
Among $22512$ events covering $18331$ hours of BiSON data from 1987 
to 1994, Chaplin \etal (1997) found $51$ events above $6.5$ times the 
mean energy, whereas less than $41$ would be expected in $90\% $ of 
the cases for an exponential distribution. Crudely speaking, about
$10$ events are unexpected in the distribution, suggesting 
$N_\XX \ge 0.05$. 

\subsection{Solar cycle variations of the total power of low degree 
p~modes \label{dis2}}

The typical energy of a low degree p mode in the range $2.7{\rm 
mHz}\le\nu\le 3.4{\rm mHz}$ considered by F98 is $<E>\sim 8\times 
10^{27}$ ergs (Chaplin \etal 1998). The velocity damping time
being $\tau_{d}\sim 3.7$ days, the flux of energy required 
to excite this p mode is $2<E>/\tau_{d}\sim 5\times 10^{22}\es$.\\
According to Libbrecht \etal (1986), the total energy in all the p modes 
(about $10^{7}$) is $10^{34}$ ergs within a factor $10$.\\
Anguera Gubau \etal (1992) and Elsworth \etal (1993) measured a 
global $30\%$ decrease of the power of low degree p modes at solar 
maximum. This decrease seems to preclude a high value of the fraction 
$\lambda$ of the total power which some additional source could 
contribute to. Nevertheless, the physical mechanisms by which the p 
mode power might
decrease, such as damping by active regions, or modification of the 
properties of the convection, have not yet been quantitatively estimated. 
This might be efficient enough to dominate the energy input due to an
additional source. Moreover, the measurement of the amplitude of global 
p modes is influenced by the presence of active regions 
covering a significant fraction of the solar surface at solar maximum 
(Cacciani \& Moretti, 1997). Given these uncertainties, no firm 
constraint can be deduced from these observations. We shall 
consider "likely" a fraction $\lambda\le 30\% $.

\subsection{Observed correlations\label{obsc}} 

F98 determined the mean correlations between the energies of low 
degree p modes at two different epochs: 160 days in 1988 near solar 
maximum using IPHIR data ($l=0$, $19\le n\le 23$ and $l=1$, $18\le 
n\le 23$), and 310 days in 1996-97 near solar minimum using GOLF data 
($l=0$ and $l=1$, $17\le n\le 25$). The mean correlation coefficient 
${\cal C}$ they measured is:
\par (i) ${\cal C}=10.7\pm 5.9\%$ in 1988 (IPHIR data), 
\par (ii) 
${\cal C}<0.6\%$ in 1996-97 (GOLF data).\\ According to F98, the 
probability that the correlation measured from IPHIR data could occur 
by chance is $0.7\%$ if the modes were independent. F98 also rejected 
the possibility of an instrumental artefact by checking that the 
noise of IPHIR data at different frequencies is not correlated.\\ 
\begin{figure}
\psfig{file=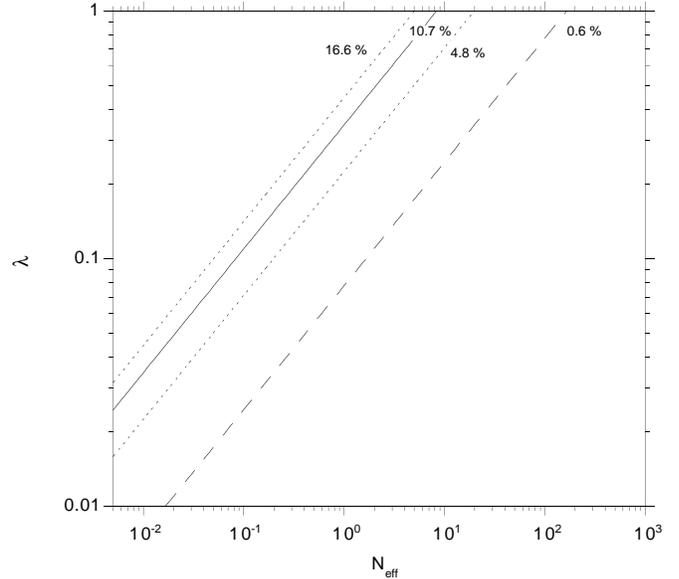,width=\columnwidth} 
\caption[]{Relationship between 
the effective number $N_{\rm eff}$ of excitations per damping time 
and the fraction $\lambda$ of the mode power due to the secondary 
mechanism of excitation, in order to match the observed correlation 
in IPHIR and GOLF periods. The value of the
correlation is indicated on each curve.}	\label{fig2}
\end{figure}
If the granules were the only source of p-mode excitation, the 
correlation would be less than $10^{-5}\%$ according to 
Eq.~(\ref{fitcor1}), \ie well below the detection limit. The actual 
limit of detection, of order $0.6\% $ (F98), corresponds to at least 
one exciting event every 29 minutes, on average.
If due to a single mechanism of excitation, the correlation measured 
with IPHIR
data would correspond to an exciting event every $12\pm 7.2$ hours.\\ 
Fig.~\ref{fig2} shows the relationship between $N_{\rm eff}$ and 
$\lambda$ required
by the correlations observed in IPHIR data, and the upper bound set 
by GOLF data.\\
The correlation measured by F98 in 160 days of IPHIR data imposes 
that $N_\XX\ge \tau_{\rm d}/160=0.02$ in the IPHIR period, which is 
coherent with the fraction of abnormal events in BiSON data.\\ 
Applying Eq.~(\ref{apllic})-(\ref{elam}) to a typical mode of energy 
$8\times 10^{27}$ ergs and damping time $\tau_{\rm d}=3.7$ days, the 
constraint ${\cal C}\ge 4.8\% $ leads us to look for a mechanism 
occurring at
most a few times per day in 1988, with an acoustic energy input of at 
least $10^{26}$ ergs per mode:
\begin{eqnarray}
0.02\le N_\XX&\le &
7.1\times \left({\beta_\XX\over4}\right) 
\left({\lambda\over0.3}\right)^{2},\\
<e_\XX>&\ge &6.7\times\left({4\over \beta_\XX}\right)
\left({0.3\over\lambda}\right)
\times 10^{26}{\rm ergs}.
\end{eqnarray}

\section{Some candidates: comet impacts, X-ray flares and coronal mass 
ejections\label{fcme}}

\begin{table}
\caption[]{Energy, momentum and occurrence rate of a free falling comet,
a X-ray flare, and a CME (Zarro \etal 1988; Crosby \etal 1993; 
Hundhausen 1997)} 
\begin{flushleft}
\begin{tabular}{cccc}
\hline\noalign{\smallskip}
& comet & flare & CME\\
\hline\noalign{\smallskip}
kinetic energy (ergs) & & &  \\
average & - & $\sim 10^{30}$ & $6.7\times 10^{30}$ \\
maximum & $2\times 10^{32}$ & $\sim 10^{33}$ & $4\times 10^{32}$\\ 
kurtosis $\beta$ & - & $\sim 33$ & 4-20 \\
momentum (g cm s$^{-1}$) & & & \\
average & - & $\sim {}10^{22}$ & $\sim 10^{23}$ \\
maximum& $6\times 10^{24}$ & - & $\sim 10^{24}$ \\
occurrence rate (day$^{-1}$) & & & \\
solar minimum & $< 0.1$ & 1 & 0.2 \\
solar maximum & $< 0.1$ & 19 & 3 \\
\hline\noalign{\smallskip}
\end{tabular}
\end{flushleft}
\label{table2}
\end{table}

\subsection{Comet impacts}

Sungrazing comets of the Kreutz group are small remnants of an 
earlier passage of a larger progenitor comet. Among the 10 sungrazers 
observed by SMM, 4 appeared during the 160 days IPHIR campaign 
(MacQueen \& St. Cyr 1991). A total of 59 sungrazers
were observed by LASCO aboard SOHO.
Even with an occurrence rate of $0.1$ per day, ${\cal C}\ge 4.8\% $ in 
Eq.~(\ref{apllic11}) would require an energy input of 
$6 \beta_\XX^{-1/2}\times 10^{27} $ ergs per event 
and per mode. Given the small scale of the comet, all $10^{7}$ p 
modes should receive the same energy input as low degree modes, which
corresponds to a total acoustic energy of $\sim 10^{34}$ ergs per 
event. This already exceeds by two orders of magnitude the kinetic 
energy of a comet with the mass as Halley's 
hitting the solar surface; not to mention the variability issue 
between IPHIR and SOHO periods, nor the difficult question of the
efficiency $f$ of the energy transfer (addressed by Gough 1994; 
Kosovichev \& Zharkova 1995).

\subsection{X-ray flares}

\subsubsection{Observations of waves generated by flares} 

Although sunspots are known to absorb high degree acoustic waves 
(Braun \etal 1987, 1988), observations of the waves generated by a
large solar flare on 24th April 1984 by Haber \etal (1988a,b) showed a 
$19\%$ increase of power in outward travelling wave, dominating the 
sunspot absorption. Nevertheless, Braun \& Duvall (1990) also 
observed the wave emission from a flare on 10th March 1989, and found 
an upper bound of $10\%$ power increase, if any.\\
Very recently, Kosovichev \& Zharkova (1998) analysed the shock wave 
produced by the "fairly moderate" flare of 9th July 1996, observed by 
MDI aboard SOHO.
The wave amplitude is associated to a momentum a factor 30 smaller 
than the one expected from their theoretical model. This unexpectedly 
high efficiency led them to conclude that the seismic flare source 
might be located in subsurface layers. \\
The effect of flares on low degree modes is however less clear.
The high energy events of BiSON data, appearing above the exponential 
energy distribution of low degree modes, do not seem to be correlated 
neither with the sunspot number, nor with the strength of X-ray flares
(Chaplin \etal 1995). Using the sum of normalized energies of low degree 
modes in IPHIR data, Gavryusev \& Gavryuseva (1997) found an 
anticorrelation between big pulses in p modes and the mean solar 
magnetic field but no correlation with the sunspot number.

\subsubsection{Theoretical estimates of p mode excitation by flares}

The theoretical estimate of the energy transmitted from a flare to p 
modes can follow three different approaches: 
\par(i) the region 
surrounding the flare is heated and expands vertically on a time scale 
short enough to communicate momentum to the atmosphere below it 
(Wolff 1972). This process could extract $10^{28}$ ergs of
acoustic energy from a $10^{32}$ ergs flare ($f\sim 10^{-4}$). 
According to Wolff (1972), the 
error bar in this estimate is at least a factor~$10$.
\par (ii) the plasma flows down towards the foot points of the field 
lines, and hits the solar surface, thus communicating momentum to it. 
This approach was followed by Kosovichev \& Zharkova (1995) who found 
a smaller effect than that observed by Haber \etal (1988a).
\par (iii) the pressure perturbation associated with the restructuring 
of the magnetic field in the flare region might be more effective, 
as suggested by Kosovichev \& Zharkova (1995). 

\subsubsection{Energy distribution of flares} 

The total energy released by a flare (less than $10^{27}$ ergs to 
$10^{33}$ ergs) is usually computed assuming that the observed hard 
X-rays are produced by
bremsstrahlung from a distribution of
accelerated electrons impinging upon a thick target plasma. Using the 
Hard X-Ray Burst Spectrometer on the Solar Maximum Mission satellite
from 1980 to 1989, Crosby \etal (1993) showed that 
the flaring rate varies by about a factor 20 between the solar 
maximum and minimum, while the energy distribution remains constant.
Applying  Eq.~(\ref{minmax}) to the correlation 
observed by IPHIR, with the $30\% $ variation of the p mode total power,
suggests the following correlation ${\cal C}_{\rm min}$ at solar minimum:
\begin{equation}
0.15\%\le{\cal C}_{\rm min}\le 0.59\%,\label{cminflare}
\end{equation}
which is compatible with the upper bound obtained from GOLF data.
\\
The occurrence rate of flares follows a power law distribution 
$p({\cal E})\sim {\cal E}^{-\gamma}$ against the total flare energy 
with a slope $\gamma<2$, indicating
that the largest flux of energy occurs in rare big energy events. Let 
${\cal E}_{\rm min}\ll {\cal E}_{\rm max}$ be the range of energies 
over which the power law distribution is observed, its mean energy 
$<{\cal E}_{\rm F}>$ and kurtosis $\beta_{\rm F}$ are: 
\begin{eqnarray}
<{\cal E}_{\rm F}>&\sim & {\gamma-1\over2-\gamma} 
\left({{\cal E}_{\rm max}\over 
{\cal E}_{\rm min}}\right)^{2-\gamma}{\cal E}_{\rm min}, 
\label{Eflare}\\
\beta_{\rm F}&\sim 
&{(2-\gamma)^{2}\over(3-\gamma)(\gamma-1)} \left({{\cal E}_{\rm max}\over 
{\cal E}_{\rm min}}\right)^{\gamma-1}. \label{sigflare}
\end{eqnarray}
According to Crosby \etal (1993), about $13$ flares 
occurred per day in 1980-1982 (solar maximum), 
with a slope of the energy distribution
$\gamma\sim1.5$ in the range $10^{28}\le {\cal E}_{\rm F}\le 10^{32}$ ergs. 
Two thirds of the flares considered by Crosby \etal (1993) fall in 
this range according to their Fig.~6, so that the occurrence rate of 
these flares can be taken as $8.7$ per day ($N_{\rm cme}\sim 32.2$).
Eqs.~(\ref{Eflare})-(\ref{sigflare}) imply $<{\cal E}_{\rm F}>\sim 10^{30}$ 
ergs and $\beta_{\rm F}\sim 33$. 
Assuming that the efficiency $f$ does not depend on the energy, the 
distributions of total energy and of energy input per mode are 
identical, and $N_{\rm eff}= 1.0$. Eqs.~(\ref{apllic}) and 
(\ref{apllic22}) 
imply:
\begin{eqnarray}
22.4\% &\le & \lambda \le 44.6\%\label{fl1}\\
<e_{\rm F}>&\ge & 10^{26} \;{\rm ergs}\;{\rm event}^{-1}\;{\rm mode}^{-1},
\label{fl2}\\
{<{\cal E}_{\rm F}>\over <e_{\rm F}>} &\le & 10^{4} \;{\rm modes}.
\label{fl3}
\end{eqnarray}
With ${\cal E}^{2}p({\cal E})\sim {\cal E}^{1/2}$, Eq.~(\ref{filtre}) 
indicates that Eq.~(\ref{fl3})
would still be correct if the energy dependence of the efficiency 
$f({\cal E})$ were to select only the most energetic events.\\
The size of the flare region viewed from earth is no more than a few 
arcminutes, so that all p modes with a degree $l\le l_{\rm max}$ 
must receive an equal amount of energy. With $l_{\rm max}=37$ according to 
Wolff (1972), this corresponds to a set of ${\cal N}\ge 10^{4}$ 
modes, which contradicts Eq.~(\ref{apllic33}).
In conclusion, the total acoustic energy input required from X-ray flares
to produce the observed correlation exceeds their total energy.

\subsection{Coronal Mass Ejections}

\subsubsection{Geometry and timing of CMEs} 

Coronal mass ejections correspond to the release of $3.3\times10^{15}$ g 
of matter on average, with a mean velocity of $350\kms$ (SMM data from 1980 
and 1984-1989, Hundhausen 1997). 
The footpoints of the outer loop are separated on average by about 45 
degrees in latitude, \ie much larger than an active region or flare 
(Harrison 1986, Hundhausen 1993). This large scale structure favours low 
degree modes.
According to Hundhausen (1994), the origin of the CME comes 
from a gradual build up and storage of energy in a pre-ejection structure 
(driven by the spreading of magnetic field lines, the emergence of 
magnetic flux, or the shear of field lines). A finite quantity of this 
energy is then released in a catastrophic breakdown in the stability 
or equilibrium of the stressed structure. When the CME is associated 
with an X-ray flare, the CME kinetic energy seems 
to be uncorrelated to the flare peak intensity (Hundhausen 1997).\\
In Hundhausen (1994), the acceleration of the CME observed by SMM on 
23rd August, 1988 was measured. Its launching precedes the flare event
(see also Harrison 1986), and the acceleration to a velocity of 
$1000$ km s$^{-1}$ occurs in less than 10 min (see the CMEs of 
17th August, 1989 and 16th February, 1986 for similar examples in 
Hundhausen 1994).\\ 
Note that a CME was associated to the flare of 9th July, 1996 analysed by 
Kosovichev \& Zharkova (1998).

\subsubsection{Solar cycle variability}

While the average mass of a CME varies little from year to year 
(Hundhausen \etal 1994b), the annual variation of their mean velocity 
does not follow the solar cycle (Hundhausen \etal 1994a). It would 
therefore be interesting to check the correlation between the high energy 
events noticed by Chaplin \etal (1995, 1997) and this new indicator.\\
According to Webb \& Howard (1994), the occurrence rate of CME varies from 
0.2 per day at solar minimum to 3 per day at solar maximum. 
As for flares, the variation of the occurrence rate is enough to 
account for the observed variation of the correlation. 
Eq.~(\ref{minmax}) implies:
\begin{equation}
0.20\%\le{\cal C}_{\rm min}\le 3.3\%.\label{cmincme}
\end{equation}

\subsubsection{Energy distribution of CMEs} 

Hundhausen \etal (1994a,b) showed that the distribution of CME velocities 
($10\kms$ to $2000\kms$) is more widely spread than their distribution of 
masses ($10^{14}$ to $10^{16}$ g). As a consequence, the
distribution of kinetic energy is spread over a much wider range than the
potential energy.\\ 
Hundhausen~(1997) selected a sample of 249 CMEs measured by SMM in 
1984-1989, associated with X-ray flares. The spread of their 
distribution of kinetic energy 
(from $10^{28}$ to $10^{33}$ ergs), measured from Fig.~1 of 
Hundhausen (1997) is such that $\beta_{\rm cme}\sim 20.2$.
Note that a smaller value ($\beta_{\rm cme}\sim 3.9$) 
is obtained when considering the distribution of squared velocities of 
$109$ CMEs during the 160 days IPHIR period in the catalogue of 
Burkepile \& St. Cyr (1993). The apparent contradiction may come 
partly from the distribution of masses, but mostly from
the high sensitivity of $\beta_{\rm cme}$ to the high energy 
events which occurred in 1989 at solar maximum. Moreover, fast CMEs are 
underrepresented in those statistics because the velocity is measured 
only when the CME is visible on more than one coronograph image 
(Hundhausen \etal 1994a).
As for flares, the distribution of acoustic energy input might be 
different from the distribution of CMEs kinetic energy, for example,
if only a subclass of CMEs excite acoustic waves. In particular,
some CMEs are known to be slowly accelerated with the solar wind, 
while others are much faster than the solar wind (MacQueen \& Fisher 
1983).\\
\begin{figure}
\psfig{file=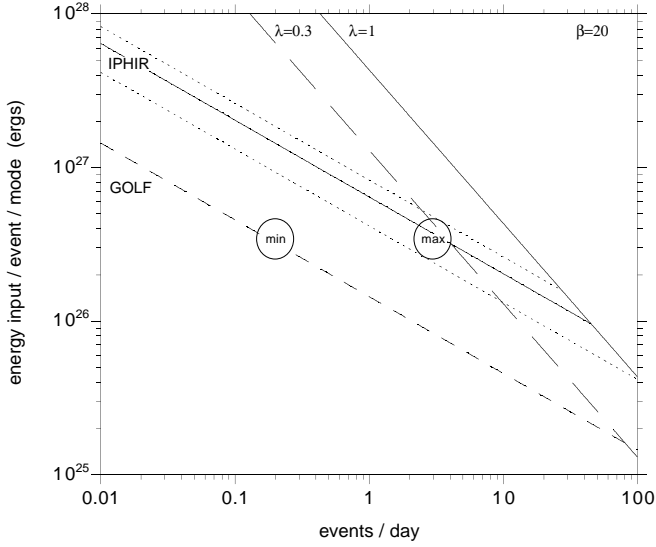,width=\columnwidth} 
\caption[]{
Relationship between the mean acoustic energy input per mode of each 
impulsive event and its occurrence rate in order to produce the correlation 
measured from IPHIR or GOLF data. The kurtosis of the
energy distribution of the source of excitation is $\beta_\XX=20$, 
the energy of the mode is $8\times 10^{27}$ ergs
and its damping time is $\tau_{\rm d}=3.7$ days. The corresponding fractions 
of total power $\lambda=1$ and $0.3$ are indicated by the continuous and 
long dashed lines. The occurrence rates of CMEs at solar minimum and maximum 
are indicated by circles.}
\label{figevent}
\end{figure}
With $\beta_{\rm cme}=20$ and an occurrence rate of $3$ CMEs 
per day ($N_{\rm cme}=11.1$, $N_{\rm eff}=0.6$), IPHIR correlations 
in Eqs.~(\ref{apllic}) and (\ref{apllic22}) imply (see 
Fig.~\ref{figevent}):
\begin{eqnarray}
 16.7\%\le \lambda  &\le& 33.2\%,\label{cme1}\\
<e_{\rm cme}>&\ge& 2.4\times 10^{26}\; 
{\rm ergs}\;{\rm event}^{-1}\;{\rm mode}^{-1},\label{cme2}\\
{<{\cal E}_{\rm cme}>\over <e_{\rm cme}>}&\le& 2.8\times 10^{4} \;
{\rm modes}.
 \label{cme3}
\end{eqnarray}
Estimates of both $\lambda$ and the acoustic energy input 
per mode $<e_{\rm cme}>$ would be multiplied by a factor 2.2 if the value  
$\beta_{\rm cme}=4$ were chosen. Although the kinetic energy
distribution of CMEs is less accurately known than for flares, one 
can infer from the published ones (Hundhausen \etal 1994b, Hundhausen 1997) 
that the product ${\cal E}^{2}p({\cal E})$ is maximum at high energy. 
As for flares, Eq.~(\ref{cme3}) would not be changed if the 
efficiency of acoustic wave generation were highest at high energy.\\
Although Eqs.~(\ref{cme1})-(\ref{cme3})
resemble Eqs.~(\ref{fl1})-(\ref{fl3}) obtained for flares, the 
following important differences should be noted:
\par(i) the estimate for CMEs is based on their kinetic energy only, 
which is certainly much smaller than the energy of the mechanism 
responsible for their ejection.
\par(ii) the large scale of CMEs might favour the excitation of low degree 
modes, especially if their ejection mechanism is rooted in the convection 
zone (${\cal N}\le 100$).
\par(iii) although the range of energy of CMEs 
(kinetic + potential $\sim 8.5\times 10^{30}$ ergs on average, 
Hundhausen \etal 1994b) is 
the same as for big flares, their momentum can be one hundred times 
larger than the momentum carried by downflowing electrons produced by 
flares (see Table~\ref{table2}). This is favourable to a higher 
efficiency of the energy transfer to acoustic modes.

\section{Conclusion}

We have investigated the consequences of interpreting the observed 
correlations in terms of an hypothetical excitation mechanism, in 
addition to the well established excitation by the granules. In particular, 
this interpretation requires impulsive events occurring no more than a 
few times per day in the IPHIR period. This has drawn our attention 
to the largest X-ray flares and CMEs at solar maximum.
The variation of their typical energy and occurrence rate 
with the solar cycle could account for the variation of the 
correlation between the IPHIR and GOLF observations.\\
If due to solar flares, the correlation determined from IPHIR data 
requires that at least $10^{-4}$ of the energy of each X-ray flares 
is injected into each low degree mode at solar maximum.
Given the number of modes (at least $10^{4}$ due to the small scale 
of flares) which should receive the same energy as low degree modes, 
we are inclined to exclude X-ray flares regardless of the efficiency 
of the acoustic wave emission by each event. This reasoning, however, 
does not exclude the possibility of 
more energetic processes associated to flares, such as the 
restructuring of the magnetic field in the flare region mentioned by 
Wolf (1972) and Kosovichev \& Zharkova (1995).\\

If the correlation is due to CMEs, at least $3.6\times 10^{-5}$ of the 
{\it kinetic} energy of each CME should be injected into each low degree 
mode at solar maximum. This leaves two possibilities open:
\par (i) CMEs may correlate only a few tens of low degree modes by 
injecting a few per cents of the CME kinetic energy into these modes. 
Higher degree modes ($l\ge 10$) would not be correlated in this case.
\par(ii) CMEs could correlate more modes if significantly more 
than the observed mechanical energy of CMEs can be extracted from 
their energy reservoir and injected into acoustic modes.\\

If all CMEs participate to acoustic emission with an efficiency $f$
independent of their energy, they should be responsible for a fraction 
$16.7\%\le \lambda\le 33.2\%$ of low degree p mode total power at 
solar maximum. Nevertheless, this fraction might be 
significantly smaller if only a subset of high energy CMEs participates 
to acoustic emission. The contribution of CMEs to the low degree p mode 
power has to be reconciled with the observed $30\%$ decrease of their 
power at solar maximum. A better theoretical understanding of the 
efficiency of the energy transfer from a CME event to acoustic modes 
is needed.\\ 

If our interpretation of IPHIR correlations is correct, we should be 
able to make the following observational tests:
\par - detect the effect of the largest CMEs on the energy of
low degree p modes, even at solar minimum,
\par - measure the correlation ${\cal C}_{\rm min}\ge 0.2 \% $ at 
solar minimum, using longer time series and more modes than F98,
\par - confirm with SOHO, during the next solar maximum, 
the correlations determined from IPHIR data, measure them with a 
better accuracy, and determine whether higher degree modes are also 
correlated.\\

Observations of the influence of CMEs on acoustic modes 
could improve our understanding of both the excitation of 
acoustic modes and the ejection mechanism of CMEs.

\acknowledgements
The author thanks T. Amari, M. Tagger, D. Gough and D. Saundby for 
helpful discussions.

\appendix

\section{Energy of a mode excited stochastically\label{app1}} 

Let us consider a spherically symmetric, adiabatic model of the sun. 
Any perturbation of velocity $v$ can be projected onto the basis of 
orthogonal
eigenfunctions ${\bf v}_{nlm}$, associated to the (supposedly real) 
eigenvalues
$\omega_{nl}$. The structure of the equations allows us to write each 
component of the velocity vector as the product of a function of $r$ 
only and a function of $\theta,\phi$ (see \eg Unno \etal 1989). This 
function is a spherical harmonic for the radial velocity $v^r$.
\begin{eqnarray}
{\bf v}({\bf r},t)&=& \sum_{nlm} A_{nlm}(t) {\bf v}_{nlm}({\bf r}),\\
{\bf v}_{nlm}({\bf r})&\equiv&\left\lbrack 
v_{nl}^r(r),v_{nl}^{h}(r){\p\over\p\theta}, 
v_{nl}^{h}(r){\p\over\sin\theta\p\phi}
\right\rbrack Y_{l}^m(\theta,\phi).
\end{eqnarray}
The eigenvectors are normalized as follows: 
\begin{eqnarray}
\int_{0}^{M_\star}|{\bf v}_{nlm}|^{2}\d M_{r}=1,\label{normv}\\ 
=\int_{0}^{R_{\star}} \rho\left\lbrack
|v_{nl}^r|^{2}+l(l+1)|v_{nl}^{h}|^{2}\right\rbrack r^{2}\d r. 
\end{eqnarray}
The spherical harmonics are written in terms of Legendre associated 
functions $P_{l}^{m}$:
\begin{eqnarray}
Y_{l}^{m}(\theta,\phi)&=&(-1)^{m+|m|\over2}q_{lm}P_{l}^{|m|}(\cos\theta) 
\e^{im\phi},\label{spha1}\\
q_{lm}&\equiv &
\left\lbrack{2l+1\over4\pi}{(l-|m|)!\over(l+|m|)!}\right\rbrack^{1\over2} .
\label{spha2}
\end{eqnarray}
The radial velocity perturbation $v_{k}({\bf r})$ is described by 
Eqs.~(\ref{defv1})-(\ref{defv3}). 
The angle $\alpha\ge 0$ made by the direction $\theta,\phi$ with 
the direction $\theta_{k},\phi_{k}$ is defined by:
\begin{eqnarray}
\cos\alpha(\theta,\theta_{k},\phi-\phi_{k})\equiv \nonumber\\
\cos\theta\cos\theta_{k}&+&\sin\theta\sin\theta_{k}\cos(\phi-\phi_{k}).
\end{eqnarray}
By projecting the perturbation onto the basis of eigenvectors, we 
obtain: 
\begin{eqnarray}
A_{nlm}(t>t_{k})&=&\cos\omega_{nl}(t-t_{k})
\int_{0}^{M_{\star}}{\bf v}_{k}({\bf 
r})\cdot{\bf v}_{nlm}^{\star}({\bf r})\d M_{r},\\
&=&2(-1)^{m+|m|\over2}a_{nlm}^{k}\e^{-im\phi_{k}}
\cos\omega_{nl}(t-t_{k}),\label{anlmreel} \\
a_{nlm}^{k}&\equiv &h_{nl}^{k}g_{lm}^{k}(\theta_{k})M_{k}^{1\over2}v_{k} 
\end{eqnarray}
where the real dimensionless functions $g_{lm}^{k}(\theta_{k})$ and 
$h_{nl}^{k}$ are defined as follows:
\begin{eqnarray}
g_{lm}^{k}(\theta_{k})&\equiv&
{q_{lm}\over 2}\int_{0}^{2\pi}\d\phi\int_{0}^{\pi}\d(\cos\theta) 
P_{l}^{|m|}(\cos\theta)\nonumber\\
&\times&\e^{-im(\phi-\phi_{k})}g_{k}(\theta,\theta_{k},\phi-\phi_{k}),\\ 
&=&q_{lm}\int_{0}^{{\alpha_{k}\over 
2}}\d\phi\int_{0}^{\pi}\d(\cos\theta) 
P_{l}^{|m|}(\cos\theta)\nonumber\\
&\times&\cos(m\phi)g_{k}(\theta,\theta_{k},\phi),\\ 
h_{nl}^k&\equiv&{1\over M_{k}^{1\over2}}
\int_{0}^{R_{\star}} \rho r^{2}v_{nl}^{r}(r)h_{k}(r)\d r. 
\end{eqnarray}
Note that the function $g_{l,m}^{k}=g_{l,-m}^{k}$ is real 
because of the cylindrical symmetry of the impulsion. \\
Eq.~(\ref{anlmreel}) corresponds to initial conditions with zero 
displacement at $t=t_{k}$, suitable to describe the transfer of impulsion 
due to a shock.\\
Assuming that the eigenfrequencies $\omega_{nl}$ are real, we add by 
hand a damping term, with a time scale $\tau_{d}$. After a series of 
excitations indexed by $k$, and neglecting the transient velocities 
present during each event, the linearity of the equations allows us 
to write the velocity as follows: 
\begin{eqnarray}
A_{nlm}(t)&=&2(-1)^{m+|m|\over2}\nonumber\\ 
&\times&\sum_{t_{k}<t}\e^{-{t-t_{k}\over\tau_{d}}}a_{nlm}^{k}
\e^{-im\phi_{k}}\cos\omega_{nl}(t-t_{k}).\label{anlmdamp} 
\end{eqnarray}
Denoting by ${\bf \xi}_{nlm}({\bf r},t)$ and ${\bf V}_{nlm}({\bf 
r},t)$ the displacement and velocity associated to the mode 
$(n,l,m)$, the acoustic energy $E_{nlm}$ of each mode is the sum of 
the kinetic and potential energies:
\begin{equation}
E_{nlm}\equiv
{1\over2}\int_{0}^{M_{\star}}(|{\bf V}_{nlm}|^{2}+ 
\omega_{nl}^{2}|{\bf \xi}_{nlm}|^{2})\d M_{r}. \end{equation}
Neglecting the slow damping compared to the fast oscillations 
($\omega_{nl}\tau_{d}\sim 7\times 10^{3}\gg 1$ for the p modes we 
consider),
and using the normalization (\ref{normv}), we may write the energy as 
follows:
\begin{equation}
E_{nlm}(t)=
{1\over2}\left(|A_{nlm}|^{2}+
{1\over \omega_{nl}^{2}}\left|{\d A_{nlm}\over \d 
t}\right|^{2}\right). \label{enlmanlm}
\end{equation}
Using Eq.~(\ref{anlmdamp}) and some algebra, we separate the two 
contributions
from the waves propagating azimuthally in opposite directions: 
\begin{equation}
E_{nlm}(t)=E_{nlm}^{+}(t)+E_{nlm}^{-}(t), 
\label{decompe}
\end{equation}
with
\begin{eqnarray}
E_{nlm}^{+}&\equiv&
\left|\sum_{t_{k}<t}
\e^{-{t-t_{k}\over\tau_{d}}}a_{nlm}^{k} 
\e^{i(\omega_{nl}t_{k}+m\phi_{k})}\right|^{2},\label{enlmdg}	\\ 
E_{nlm}^{-}&\equiv&E_{nl-m}^{+}.
\end{eqnarray}
Although Eqs.~(\ref{anlmdamp}) and (\ref{enlmanlm}) imply 
$E_{nlm}=E_{nl-m}$, the two components $E_{nlm}^{+}$ and 
$E_{nlm}^{-}$, are not
equal, except on average.\\
This separation of the components of the energy is important since 
the rotation enables us to separate these two components. Let us restrict 
ourselves to the simplest case of a solid body rotation, and neglect 
the Coriolis forces. This is equivalent to
replacing in Eq.~(\ref{enlmdg}) the azimuthal angle $\phi$ by 
$\phi+\Omega t$ and $\phi_{k}$ by $\phi_{k}+\Omega t_{k}$, and thus 
we obtain Eqs.~(\ref{enerstoc})-(\ref{defphi}). $E_{nlm}^{+}$ is the 
energy associated to the frequency $\omega_{nl}+m\Omega$, while 
$E_{nlm}^{-}$ is associated to the frequency $\omega_{nl}-m\Omega$.\\
The average energy input $<e_{nlm}>$ of a single random excitation onto 
the mode $n,l,m$ is proportional to the total kinetic energy 
${\cal E}$ of the perturbation: 
\begin{equation}
<e_{nlm}>\equiv <2h_{nl}^{2}g_{lm}^{2}{\cal E}>
.\label{squav} 
\end{equation}

\section{Correlation produced by a single excitation mechanism}

\subsection{Case $N\ll1$}

In order to treat the general case, we need to establish first some 
properties of the case $N\ll1$. The index $nlm$ of the energy 
is omitted in what follows, for the sake of clarity
The mean energy $<E>$ of the damped oscillator is directly 
proportional to the mean energy input $<e>$ of each impulsive event:
\begin{eqnarray}
{1\over\Delta t}\int_{0}^{\Delta t}(\e^{-{t\over\tau_{\rm d}}})^{2}\d 
t&\sim&{N\over2},\\
<E>&\sim&{N\over 2}<e>\label{meanE},\\
<E^{2}>&\sim&{N\over 4}<e^{2}>.\label{meanE2}
\end{eqnarray}
The energies $E,E'$ of two modes $nlm$, $n'l'm'$ satisfy:
\begin{equation}
{<EE'>\over <E><E'>}\sim {1\over N}
{<ee'>\over <e><e'>}.
\label{meanEE}
\end{equation}

\subsection{General case}

The correlation in the general case is derived using the following 
theorem: if a continuous function $f(x)$ satisfies $f(2x)=f(x)$ for 
any value of $x$, and $\lim_{x\to0}f(x)$ exists, then $f$ is 
constant.\\
Let us divide the random series of excitation into two independent 
subset of same statistical characteristics, with an occurrence rate $N/2$. 
Denoting by $E(N/2)$ the energy of the mode excited by only one 
subset, one can check from Eq.~(\ref{enerstoc}) that: 
\begin{equation}
<E(N)>=2<E(N/2)>,
\end{equation}
and conclude that $f(N)\equiv E/N$ is a constant, which we deduce from
the case $N\ll1$ (Eq.~\ref{meanE}):
\begin{equation}
<E>\equiv {N\over2}<e>.
\label{energen}
\end{equation}
The variance of the energy can also be written as follows:
\begin{equation}
{\Var E\over <E>^{2}}(N)={1\over 2}{\Var E\over <E>^{2}}(N/2)+{1\over 2}.
\end{equation}
This implies that $f(N)\equiv N(\Var E/<E>^{2} -1)$ is 
a constant, which we deduce from the case $N\ll1$ (Eq.~\ref{meanE2}):
\begin{equation}
{\Var E\over <E>^{2}}=
1+{1\over N}{<e^{2}>\over <e>^{2}}.
\label{vaava}
\end{equation}
The correlation ${\cal C}$ between two modes $nlm$, $n'l'm'$ is 
defined as:
\begin{equation}
{\cal C}\equiv {<EE'>-<E><E'>\over 
(\Var E)^{1\over2}(\Var E')^{1\over2}}.\label{defcor}
\end{equation}
Using the fact that
\begin{equation}
{<EE'>\over <E><E'>}(N)={1\over2}{<EE'>\over <E><E'>}(N/2)+{1\over2},
\end{equation}
we conclude that $f(N)\equiv N(<EE'>/(<E><E'>)-1)$ is a constant, which we 
deduce from the case $N\ll1$ (Eq.~\ref{meanEE}). The correlation expressed 
by Eq.~(\ref{corgen1}) is then derived from 
Eqs.~(\ref{vaava})-(\ref{defcor}).

\section{Correlation produced by two excitation mechanisms\label{Aseveral}}

Let us consider two independent series of impulsive events due to two 
different excitation mechanisms superimposed, indexed by $k$, 
characterized by the series of velocity perturbations 
$v_{k}^{(1)},v_{k}^{(2)}$, and mean interpulse times 
$\Delta t^{(1)},\Delta t^{(2)}$.
The energy of a mode $(nlm)$ can be decomposed into two random walks 
of $N_{1}$ and $N_{2}$ steps.
\begin{eqnarray}
E_{nlm}^{+}(t)=\nonumber\\ 
\left | 
\sum_{t_{k}^{(1)}<t} \e^{-{t-t_{k}^{(1)}\over\tau_{d}}}
a_{nlm}^{k(1)}\e^{i\Phi_{nlm}^{k(1)}}
+
\sum_{t_{k}^{(2)}<t} \e^{-{t-t_{k}^{(2)}\over\tau_{d}}}
a_{nlm}^{k(2)}\e^{i\Phi_{nlm}^{k(2)}} 
\right |^{2}.
\end{eqnarray}
Let us write this equation for two modes $(nlm)$ and $(n'l'm')$. 
Hereafter we simply denote the quantities specific to the mode 
$(n'l'm')$ with a prime, for the sake of clarity. $E_{i}$ and $E'_{i}$ 
are the energies of the modes $nlm$ and $n'l'm'$ excited by the
excitation mechanism $(i)$ only ($i=1,2$).
The series of phases 
$\Phi^{k(1)},\Phi^{k(2)},\Phi^{k(1)'},\Phi^{k(2)'}$ are independent.
\begin{eqnarray}
<E>&=&<E_{1}>+<E_{2}>,\\
{\rm Var}(E)&=&{\rm Var}(E_{1})+{\rm Var}(E_{2})+2<E_{1}><E_{2}>,
\label{varsom}\\ 
{\rm Var}(E)\;{\cal C}&=&{\rm Var}(E_{1}){\cal C}_{1}+ 
{\rm Var}(E_{2}){\cal C}_{2}\label{corsom},
\end{eqnarray}
where ${\cal C}_{i}$ is the correlation between $E_{i}$ and $E'_{i}$. 
Let us define the fractions  $\lambda,\lambda'$ of the total 
power of each mode contributed by the second mechanism as follows: 
\begin{eqnarray}
\lambda&\equiv& {<E_{2}>\over <E>},\\
\lambda'&\equiv &{<E_{2}>\over <E>}.
\end{eqnarray}
Let convection be the first excitation mechanism ($C_{1}= 0$), and let
us use the index $({\XX})$ for the properties $N_{\XX},e_{\XX}$ of the 
second excitation mechanism. Using Eqs.~(\ref{vaava}), (\ref{corgen1}) 
and (\ref{varsom}) in (\ref{corsom}), we obtain:
\begin{eqnarray}
{\cal C}={<ee'>\over
<e^{2}>^{1\over2}<e'^{2}>^{1\over2}}\nonumber\\
\times\left(
1+{N_{\XX}\over\lambda^{2}}
{<e>^{2}\over<e^{2}>}\right)^{-{1\over2}}
\left(
1+{N_{\XX}\over\lambda'^{2}}
{<e'>^{2}\over<e'^{2}>}\right)^{-{1\over2}}.
\label{corgenlamb}
\end{eqnarray}
Let $p_{\rm T}({\cal E})$ be the density of probability of the additional 
source of energy as defined in Sect.~\ref{effic}. 
The number of efficient excitations per damping time is independent 
of $f_{\XX}$:
\begin{equation}
N_\XX=N_{\rm T}\int_{{\cal E}_{1}}^{{\cal E}_{2}}
p_{\rm T}({\cal E})
\d {\cal E}.\label{ap1}
\end{equation}
The mean energy and kurtosis of the distribution of efficient events are:
\begin{eqnarray}
<{\cal E}_\XX>&=& f_{\XX}
{\int_{{\cal E}_{1}}^{{\cal E}_{2}}{\cal E}p_{\rm T}({\cal E})
\d {\cal E}
\over
\int_{{\cal E}_{1}}^{{\cal E}_{2}}p_{\rm T}({\cal E})\d {\cal E}},
\label{ap2}\\
\beta_\XX&=&{\int_{{\cal E}_{1}}^{{\cal E}_{2}}p_{\rm T}({\cal E})
\d {\cal E}
\int_{{\cal E}_{1}}^{{\cal E}_{2}}{\cal E}^{2}p_{\rm T}({\cal E})
\d {\cal E}
\over 
\left(\int_{{\cal E}_{1}}^{{\cal E}_{2}}{\cal E}p_{\rm T}({\cal E})
\d {\cal E}
\right)^{2}}
.\label{ap3}
\end{eqnarray}
The kurtosis is also independent of $f_{\XX}$.
Eq.~(\ref{filtre}) is obtained by incorporating 
Eqs.~(\ref{ap1})-(\ref{ap3}) into Eq.~(\ref{apllic11}).

\end{document}